\documentclass[preprint,5p,twocolumn]{elsarticle}

\usepackage{hyperref}

\usepackage{amsmath}
\usepackage{amssymb}
\usepackage{amsthm}
\usepackage{mathtools}
\usepackage{braket}
\usepackage{subfig}

\newcommand{\numberset}{\mathbb}
\newcommand{\R}{\numberset{R}}
\newcommand{\Z}{\numberset{Z}}
\usepackage{graphicx}
\usepackage{dcolumn}
\usepackage{bm}
\usepackage{graphicx}  
\usepackage{dcolumn}   
\usepackage{bm}        
\usepackage{amssymb}   
\usepackage{setspace}
\usepackage{hyperref}
\usepackage{textcomp}
\usepackage{amsmath}
\usepackage{palatino}
\usepackage{amsfonts}
\usepackage{widetext}

\usepackage{cuted}

\journal{Journal of \LaTeX\ Templates}

\begin{document}

\begin{frontmatter}

\title{Bianchi I model as a prototype for a cyclical Universe}

\author{Andrea Marchi\fnref{cor01}}
\ead{afmanga2000@gmail.com}
\address{Dipartimento di Fisica (VEF), P.le A. Moro 5 (00185) Roma, Italy,}

\author{Giovanni Montani}
\ead{giovanni.montani@frascati.enea.it}
\address{Dipartimento di Fisica (VEF), P.le A. Moro 5 (00185) Roma, Italy, and \\
ENEA, Unità Tecnica Fusione, ENEA C. R. Frascati, via E. Fermi 45, 00044 Frascati (Roma), Italy,}

\author{Riccardo Moriconi}
\ead{moriconi@na.infn.it}
\address{Dipartimento di  Fisica "E. Pancini", Universit\`{a} di Napoli "Federico II",
 Compl. Univ. di Monte S. Angelo, Edificio G, Via Cinthia, I-80126, Napoli, Italy, and \\
 Istituto Nazionale di  Fisica Nucleare (INFN) Sez. di Napoli, Compl. Univ. di Monte S. Angelo,
Edificio G, Via Cinthia, I-80126, Napoli, Italy,}

\fntext[cor01]{Corresponding author}




\begin{abstract}
We analyze the dynamics of the Bianchi I model in the 
presence of stiff matter, an ultrarelativistic component and 
a small negative cosmological constant. We quantize this model in the framework of the polymer quantum mechanics, in order to introduce cut-off 
features in the minisuperspace dynamics. 

We then apply to the polymer Wheeler-DeWitt equation, emerging from the Dirac constraint, an adiabatic approximation a la Vilenkin, 
which treats the Universe volume as a quasi-classical variable, 
becoming de facto the dynamical clock for the pure quantum degrees of freedom, here identified in the Universe anisotropies. 

The main issue of the present analysis consists of 
determining a cyclical evolution for the Bianchi I model, 
oscillating between the Big-Bounce induced by the cut-off physics 
and the turning point due to the small cosmological constant. 
Furthermore, the mean value of the Universe anisotropy variables remains finite during the whole evolution, including the phase across the 
Big-Bounce. Such a feature, according to a suitable choice of the initial conditions makes the present cosmological paradigm, 
a viable scenario for the description of a possible primordial and late phases of the actual Universe. 
\end{abstract}

\begin{keyword}
Quantum cosmology \sep Bianchi Model \sep Polymer quantum mechanics
\MSC[2010] 00-01\sep  99-00 
\end{keyword}

\end{frontmatter}

\section{Introduction}

The existence of a singularity in the 
dynamics of the cosmological models is a 
very general feature of the Einstein equations, 
see \cite{BKL82} for a characterization 
of the generic cosmological solution and 
\cite{hawking} for general theorems 
on this topic.

The idea that the non-physical feature of 
the singularity in the cosmological problem could be removed by quantum
effects has been reliably proposed since very long time \cite{primordial}. 
However, the implementation of the canonical quantum gravity, i.e. of the
so-called Wheeler-DeWitt equation
\cite{DeWitt1},\cite{DeWitt2},\cite{DeWitt3}, did not provide a clear
answer in this 
direction and, in general, the singular point 
remains dynamically available, especially in the sense of a quasi-classical
limit 
of the quantum theory. 
In particular, in a general minisuperspace representation, the volume of the Universe takes the role of the time
variable for the quantum dynamics and therefore 
its vanishing value (corresponding to the energy density divergence) is always accessible for the system evolution 
\cite{misner},\cite{misnermixmaster},\cite{QuaMontani}. 

Clearly, approaches of canonical quantum gravity  to specific models (see
\cite{moriconiBB1},\cite{pittorino}) of different quantum gravity
approaches 
\cite{veneziano},\cite{barrow}, 
were able to provide non-singular and, 
in case, cyclical Universes. 

However, the idea that the so-called 
Big-Bounce could replace the Big-Bang singularity emerged in a more general
perspective when it became clear that the implementation of Loop Quantum
Gravity \cite{QuaMontani},\cite{rovelli} to the 
cosmological problem could 
provide a reliable cut-off on the physics of the singularity.  
For a review of the most significant results obtained in this field about
ten/fifteen years ago, see \cite{loop1},\cite{loop2},\cite{loop3}. 
Despite this Loop quantum Gravity Theory 
is not at all free from criticisms 
\cite{cianfrani1},\cite{cianfrani2},\cite{cianfrani3} and 
applicability limits to more general models than the simplest isotropic
Universe 
\cite{boj1},\cite{ashBKL}, nonetheless, from 
the derivation of the Big-Bounce in the Loop Quantum Cosmology, 
a new point of view emerged in theoretical cosmology, based on  the idea that
a 
cut-off on the singularity could naturally appear as effect of a
non-perturbative treatment of the quantum gravity effects. 

An interesting procedure of quantization, able to account the discreteness of
the configurational variables, is provided by the so-called polymer quantum mechanics\cite{corichi},\cite{corichi2}. 
The implementation of this revised quantum mechanical approach to the
minisuperspace variables, 
in particular to the Universe volume, 
demonstrated its capability to mimic some important features of the Loop
Quantum Cosmology, but the possibility to deal with a simpler metric approach,
makes the polymer minisuperspace a much more viable procedure, even for rather
complicated cosmological models, with respect to the cosmological restriction of the 
Loop Quantum gravity. 
For a recent discussion of the Big-Bounce cosmology, associated  to the
dynamics of the Bianchi I and IX model in Polymer quantum Cosmology, see \cite{moriconiBB2}. 

Here, we complete and generalize the ideas proposed 
in that paper, by considering the polymer quantum dynamics of a Bianchi type I model, containing three different matter contributions: 
stiff matter, ultrarelativistic matter and a small negative 
cosmological constant (the latter has been implemented in 
a different context in \cite{moriconiBB1}. 

The dynamics of the model is described by adopting an 
adiabatic separation of the polymer quantum dynamics, 
according to the Vilenkin formulation for the Universe wavefunction interpretation \cite{vilenkin}. 

The variable we consider as the semi-classical one is the 
Universe volume, i.e. a function of the isotropic cosmic scale factor, in terms of which also the matter contributions can be expressed. The anisotropy variables, described via the standard Misner variables 
$\beta_{\pm}$, remain instead real quantum degrees of freedom and their dynamics is analyzed via the Ehrenfest theorem, also 
compared to the wave packet behavior.

By this scenario, we are able to demonstrate that the Bianchi I 
dynamics is reduced to a cyclical Universe, oscillating between the bounce, ensured via the polymer cut-off physics and the later turning point due to the negative cosmological constant. 
Furthermore, we are able to follow the dynamics of the Universe anisotropies, 
showing how their values remain finite during all the evolution, even across the bounce point. This last result is relevant in view 
of implementing in the thermal history of the present Universe a bounce cosmology 
\cite{brandenberger}, since it is always possible 
to remain sufficiently close to an isotropic Universe. 

The main merit of the present analysis is to clarify how 
the dynamics of an intrinsically anisotropic model, 
like the type I, can be reconciled with a possible scenario 
for the present Universe, as soon as cut-off and semi-classical effects are considered in its morphology. 
However, more than to link the actual model with the 
thermal history of the present Universe, here we aim to 
calculate the anisotropy behavior across the bounce and 
their later evolution, in order to clarify how they no longer explode 
and can be maintained under control in the Universe dynamics, via suitable initial conditions. 

The paper is structured as follows.

In Section \ref{sec:1} we introduce the basic cosmological features related to the flat homogeneous and isotropic Friedmann-Roberston-Walker model and to the to the flat homogeneous one, i.e. the Bianchi I model.

Then, in Section \ref{sec:2}, it is taken in consideration the polymer representation of the quantum mechanics. In particular, for the one-dimensional particle case, we introduce the kinematic and the dynamical proprieties of this approach in order to analyze the modification induces by the polymer implementation of the quantum mechanics.

The Section \ref{sec:3} is dedicated to the introduction of the Vilenkin approach for the wave function of the Universe. In this scheme, here reported as in the original Vilenkin work, it is considered, for a generic homogeneous universe, the semiclassical dynamics (by studying the Hamilton-Jacobi equation related to the classical action) and the evolution of the variables related to the behavior of the pure quantum degrees of freedom of the system.

Then, in Section \ref{sec:4}, the Bianchi I model in presence of several kind of matters is considered in the scheme of the Misner-like variables. In particular, by choosing the Vilenkin interpretation for the wave function of the Universe, we analyze from the semiclassical point of view the evolution of the isotropic variable (related to the volume of the Universe) towards the singularity and from the quantum point of view the behavior of the quantum degrees of freedom: the anisotropies.

Moreover, the polymer generalization of the previous described model is illustrated in Section \ref{sec:5}. In particular, we consider the modifications induced in the configuration variables dynamics towards the initial singularity when a polymer discretization of the Universe volume occurs.

A section dedicated to concluding remarks complete the paper.

\section{Cosmological Framework}
\label{sec:1}
In this section, we provide the basic  cosmological paradigm to  
properly interpret our analysis, with particular reference to the 
dynamical characterization of the considered model. 
We start by recalling the basic evolution scheme of the 
isotropic Universe, associated to the Robertson-Walker geometry, 
in order to provide a clear link to the adopted Bianchi I model, 
which corresponds to a generalization of the isotropic flat Universe. 
Then, we will give the full dynamical scheme, for what concerns both the geometry and the matter sources, on which the set up of 
the Bianchi I model as a cyclical Universe is based. 

\subsection{Isotropic Universe}
When we study the Universe, we can see that on a scale large enough ($\gtrsim 100Mpc$)  the spatial
distribution of matter in the universe is homogeneous and isotropic, as stated by the \textit{Cosmological Principle}\cite{kolb}. In a
synchronous frame, the metric that corresponds to such a space-time is the Robertson-Walker(RW) geometry, whose line element is
\begin{equation}
\label{eqn:RWline}
    ds^2 = -c^2 dt^2 +a(t)^2 dl^2,
\end{equation}
\begin{equation}
\label{eqn:dl}
dl^2 = \frac{1}{1-k r^2}dr^2 + r^2 d\theta ^2 + r^2 \sin^2 \theta d \phi^2
\end{equation}
where the coordinates $(t,r,\theta,\phi)$ are comoving coordinates, $a(t)$ is the cosmic scale factor and $k$, after a proper redefinition of the scale factor, can take only the values $\{+1,-1,0 \}$ if the  space is respectively at constant positive, negative, or zero spatial curvature.

In order to obtain information about the dynamics of the homogeneous and isotropic Universe, we have to study the Einstein equations

\begin{equation}
\label{eqn:Einstein}
G^{\nu}_{\mu} = \chi T^{\nu}_{\mu},
\end{equation}
where $\chi = \frac{c^4}{16 \pi G} $ and $T^{\nu}_{\mu}$ is the energy-momentum tensor. 

The requirement for the stress-energy tensor $T_{\mu \nu}$ is to be consistent with the symmetries of the metric (homogeneity and isotropy). For this reason, it would be necessary diagonal and with equal spatial components. The simplest fulfillment of this requirements is the perfect fluid case, with energy density $\rho(t)$ and pressure $p(t)$:
\begin{equation}
\label{SET}
T_{\mu \nu} = diag (\rho, -p,-p,-p).
\end{equation}

Therefore, if we take the 0-0 component of the Eq. (\ref{eqn:Einstein}) for the line element (\ref{eqn:RWline}) we obtain the well-known Friedmann equation for the isotropic Universe
\begin{equation}
\label{eqn:friedmann}
    H^2(t) \equiv \left(\frac{\dot{a}}{a}\right)^2 = \frac{8 \pi G}{3 c^2}\rho - \frac{k c^2}{a^2}.
\end{equation}
Furthermore, making the difference between the $i-i$ component and the Eq.(\ref{eqn:friedmann}) we lead to the the acceleration equation
\begin{equation}
\label{eqn:acceleration}
\frac{\ddot{a}}{a} = - \frac{4 \pi G}{3 c^2}(\rho + 3p)
\end{equation}
where we have a decelerating Universe when $p > -\frac{1}{3}\rho$. From thermodynamic
considerations we obtain a continuity equation that give us an expression for the energy density as a function of the scale factor
\begin{equation}
\label{eqn:rhoofa}
\rho (a) = \frac{\mu^2}{a^{3(1+w)}}
\end{equation}
where $\mu$ is a constant parameter and $w$ came from the equation of state $p=w \rho$. It is also possible to
obtain, for different values of the parameter of state $w\simeq const.$, the explicit dependence of the scale factor as a function of the synchronous time. Hence, for the case $k=0$ or $a \to 0$ from the Friedmann equation we get
\begin{equation}
\label{eqn:aoft}
a(t) \propto t^{\frac{2}{3(w+1)}}.
\end{equation}

In this paper, we will adopt a generalization of the present cosmological paradigm in order to account for Universe's anisotropies, still preserving the homogeneity.

\subsection{Bianchi I dynamics}
When we consider only homogeneous Universe with different anisotropic degree we obtain Bianchi's Universe, that are a classification of tridimensional Lie's group calculated
starting from the structure constant
\begin{equation}
\label{eqn:bianchiconstant}
C^c_{ab}= \biggl ( \frac{\partial e^c_{\alpha}}{\partial x^{\beta}} - \frac{\partial e^c_{\beta}}{\partial x^{\alpha}} \biggr ) e^{\alpha}_a e^{\beta}_b
\end{equation}
where $e^a$, $e_a$ are cartesian versor with component $e^a_{\alpha}$, $e^{\alpha}_a$.
In an homogeneous spacetime, such a constants can be naturally rewritten as
\begin{equation}
\label{eqn:bianchiconstant2}
C^{ab} = n^{ab} + \varepsilon^{abc} a_c
\end{equation}
where $n^{ab} = n^{ba}$ is a symmetric matrix, $a_a = C^{b}_{ba}$ is a dual vector
and $\varepsilon_{abc} = \varepsilon^{abc}$ is the totally antisymmetric tridimensional
Levi-Civita tensor. Through Jacobi Identities 
\begin{equation}
\label{eqn:jacobi}
\bigl[ [e_a,e_b],e_c \bigr] + \bigl[ [e_b,e_c],e_a \bigr] + \bigl[[e_c,e_a],e_b \bigr] = 0
\end{equation}
we obtain $n^{ab} a_b= 0$. Now, if we redefine $a_b$ vector as $a_b = (a,0,0)$ and
$n^{ab}$ matrix as $n^{ab} = diag\{ n_1, n_2, n_3\}$, we can rewrite~\eqref{eqn:bianchiconstant2} as $an_1 = 0$, whose solutions are $a=0$ or $n_1 = 0$. When $a=0$ we have six Class $A$ Models and when $a \neq 0$ we have three Class $B$
Models. In the next, we will focus on the Bianchi I model, the simplest one, for which
$a=0$ and $n_1=0$, \textit{i.e} the constant of structures (and the spatial curvature) vanishes. The Bianchi I model is the natural anisotropic extension of FRW model
with $k=0$, so it generalizes an homogeneous flat Universe. 

The description
of the model will be done with respect to the Misner-like variables 
$ \{ a, \beta_{\pm} \}$(the original isotropic Misner variable is $\alpha = \ln a$).

Following the Misner parametrization\cite{misnermixmaster}, the line element for a generic Bianchi model can be written as
\begin{equation}
\label{pappa}
ds^{2} = N(t)^{2}dt^{2} - \eta_{ab}\omega^{a}\omega^{b},
\end{equation}
where 
$\omega ^{a} = \omega^{a}_{\alpha}dx^{\alpha}$ is a set of three invariant differential forms, $N(t)$ is the lapse function ($N(t)=1$ in synchronous time) and $\eta _{ab}$ is defined as $\eta _{ab} = a^{2}(e^{2\beta})_{ab}$. 
In this scheme, $a$ expresses the isotropic volume of the universe (for $a\rightarrow 0$, the initial singularity is reached.), while the matrix $\beta_{ab} = diag(\beta _{+} + \sqrt{3}\beta _{-} , \beta _{+} - \sqrt{3}\beta _{-} , -2\beta _{+})$ accounts for the anisotropy of this model.

All the dynamical information about the Bianchi I model in the vacuum case is collected, in the Arnowitt-Deser-Misner (ADM) formalism, in the associated superHamiltonian constraint $\mathcal{H}$, that in the Misner-like variables takes the form
\begin{equation}
\label{eqn:hbianchinomatter}
N(t)\mathcal{H} = \frac{G N(t)}{24 \pi a^3 c^3} \bigl[ -a^2 p^2_a + p^2_{+} + p^2_{-} \bigr],
\end{equation}
 and $ \{ p_{a},p_{+},p_{-} \} $ are the conjugated momenta related to the Misner-like variables. Through Eq.(\ref{eqn:rhoofa}) it is possible to
introduce a matter contribution inside the superHamiltonian when the
opportune $w$ is choosen. Indeed, the expression of the energy density depends only on the volume of the Universe (\textit{i.e.} $a^{3}$) and not on the anisotropic variables $\{ \beta_{+}, \beta_{-} \}$, as confirmed by the conservation law by the energy-momentum tensor $\nabla^{\mu}T_{\mu \nu} = 0$\cite{gravitation}.

In Section \ref{sec:4} we will take three different contribution, given by the stiff matter ($w=1$), the negative cosmological constant ($w=-1$) and the ultra relativistic case (radiation)
($w=1/3$).

\section{Polymer Quantum Dynamics}
\label{sec:2}

Now we summarize the fundamental features of the polymer
quantization scheme and we will give a general picture of the model through the kynematical and dynamical properties. The Polymer representation of quantum mechanics is non-equivalent to the usual Schr\"{o}dinger quantum mechanics approach, and it is based on a modified  Commutation Rules and it represents a useful tool when one or more of the phase space variables are discretized\cite{corichi}. 
Let us consider a set of kets $\ket{\mu_i}$ with $\mu_i \in \R$ and discrete index $i= 1, \dots , N$. The vectors $\ket{\mu_i}$ belong to the Hilbert space $\mathcal{H}_{poly} = L^2 ( \R_b , d \mu_H )$. The inner product between two kets is $\braket{\nu|\mu} = \delta_{\nu, \mu}$ and system state is described by a generic linear combination of them
\begin{equation}
\label{eqn:polymerstate}
\ket{\psi} = \sum^N_{i=1} a_i \ket{\mu_i}
\end{equation}
We can identify two fundamental operators in this space, label operator $\hat{\varepsilon}$ and shift operator $\hat{s}(\lambda)$, that act as follows
\begin{equation}
\label{eqn:labelshift}
\hat{\varepsilon}\ket{\mu} = \mu\ket{\mu} \quad, \quad \hat{s}(\lambda)\ket{\mu} = \ket{\mu + \lambda}
\end{equation}
where the parameter $\lambda$ is arranged in the quantization procedure but, in principle, it can takes all the real values on the positive axes.
This one-dimensional system is characterized by a the position variable $q$ and the conjugate momenta $p$.

In the next we will make the choice to assign a discrete characterization to the variable $q$ and to describe the wave function of the system in the $p$-polarization. The projection of the states on the pertinent basis vectors is
\begin{equation}
\label{eqn:pprojection}
\phi_\mu(p) = \braket{p|\mu} = e^{\frac{i}{\hbar}\mu p}
\end{equation}
and it is possible to show that the label operator represents exactly the position operator after the introduction of two unitary operators $U(\alpha) = e^{i \alpha \hat{q}}$ and $V(\beta) = e^{i \beta \hat{p}}$, $(\alpha, \beta) \in \R$ which obey the Weyl Commutation Rules (WCR) $U(\alpha) V(\beta) = e^{i \alpha \beta} V(\beta) U(\alpha)$. However, it is not possible to define a differential momentum operator, as a consequence of the discontinuity for $\hat{s}(\lambda)$ pointed out in~\eqref{eqn:labelshift}.

To study the dynamical properties of this model we have to investigate the system from the Hamiltonian point of view. A one-dimensional particle of mass $m$ in a potential $V(q)$ is describing by the Hamiltonian
\begin{equation}
\label{eqn:hamiltonian}
H = \frac{p^2}{2m}+V(q).
\end{equation}
Being $q$ a discrete variable, it is impossible to define, in the $p$-polarization, the operator $\hat{p}$ as a differential operator. To solve this problem we define a subspace $\mathcal{H}_{\gamma_\lambda}$ of $\mathcal{H}_{poly}$. This subspace contains all vectors that live on the lattice of points identified by the lattice spacing $\lambda$ in this way
\begin{equation}
\label{eqn:lattspace}
\gamma_\lambda = \{ q \in \R | q = n\lambda, \forall n \in \Z \},
\end{equation}
where $\lambda$ has the dimensions of a length. This means that the basic vectors take the form $\ket{\mu_\lambda}$ with $\mu_\lambda = \lambda n $ and the states are definied as a linear combination of them
\begin{equation}
\label{eqn:polystate}
\ket{\psi}=\sum_n b_n \ket{\mu_n}.
\end{equation}
The basic procedure of the polymer quantization consists in the approximation of the term corresponding to $\hat{p}$, the non-existing operator, and to find for this term an appropriate and well-defined quantum operator. In both $p$ and $q$ polarizations, the operator $\hat{V}$ is exactly the shift operator $\hat{s}$ and, through this identification, it is possible to write an approximate version of $\hat{p}$. Thus, for $ p \ll \hbar/\lambda$, we get
\begin{equation}
\label{eqn:papprox}
p \simeq \frac{\hbar}{\lambda}\sin{ \biggl( \frac{\lambda p}{\hbar}\biggr)} = \frac{\hbar}{2 i \lambda} \biggl( e^{\frac{i}{\hbar}\lambda p} - e^{- \frac{i}{\hbar}\lambda p} \biggr),
\end{equation}
and we can now define a new regulated operator $\hat{p}_\lambda$ as
\begin{equation}
\label{eqn:hatp}
\hat{p}_\lambda \ket{\mu_n} = \frac{\hbar}{2 i \lambda}\bigl ( \ket{\mu_{n-1}} - \ket{\mu_{n+1}}\bigr ).
\end{equation}
It is also possible to define an approximate version of $\hat{p}^2$, always for $p \ll \hbar/\lambda$ in an analogous way:
\begin{equation}
\label{eqn:p2approx}
p^2 \simeq \frac{2 \hbar^2}{\lambda ^2} \biggl[ 1 - \cos{ \biggl ( \frac{\lambda p}{\hbar}\biggr )}\biggr] = \frac{2 \hbar^2}{\lambda^2} \bigl[ 1 - e^{\frac{i}{\hbar}\lambda p} - e^{-\frac{i}{\hbar}\lambda p}\bigr ],
\end{equation}
whose operatorial version is
\begin{equation}
\label{eqn:hatp2}
\hat{p}^2_\lambda \ket{\mu_n} = \frac{\hbar^2}{\lambda ^2} \bigl[ 2\ket{\mu_n} - \ket{\mu_{n+1}} - \ket{\mu_{n-1}} \bigr].
\end{equation}
Remembering that $\hat{q}$ is a well-defined operator as in the canonical way (\textit{i.e.} $\widehat{q} = i\hbar \frac{d}{d p}$), the approximate version of the starting Hamiltonian~\eqref{eqn:hamiltonian} is
\begin{equation}
\hat{H}_\lambda = \frac{1}{2m}\hat{p}^2_\lambda + V(\hat{q})
\end{equation}
The Hamiltonian operator $\hat{H}_\lambda$ is a well-defined and symmetric operator belonging to $\mathcal{H}_{\gamma_\lambda}$.

This revised procedure of quantization is particular relevant when applied to the scale factor variables, since it becomes a tool to account for cut-off physics in the Universe quantum dynamics.

\section{Vilenkin}
\label{sec:3}
In quantum cosmology the Universe is described by a wave function, which represents the solution of the Wheeler-deWitt (WDW) equation. Let us consider the homogenous minisuperspace model\cite{primordial}. 

The action for this class of model is
\begin{equation}
    \label{eqn:vilaction}
    S = \int dt [p_\alpha \dot{h}^\alpha - N (g^{\alpha \beta} p_{\alpha} p_{\beta} + U(h))],
\end{equation}
where $h^\alpha$ represents the superspace variables, $p_\alpha$ are the conjugated momenta to $h^\alpha$, $N=N(t)$ is the lapse function, $g^{\alpha \beta}$ is the superspace metric and $U(h)$ takes into account the spatial curvature and the potential energy of the matter field. In order to achieve a clear probabilistic interpretation of the Universe wabe function, we
follow the Vilenkin approach\cite{vilenkin} in which there is a separation of the configuration variables in two classes: semiclassical and quantum ones. This way the quantum variables represent a small subsystem of the Universe and the effects of this variables on the semiclassical ones are negligible. We will use $h_\alpha$ to denote semiclassical variables and $q^\nu$ the quantum variables. The WDW equation for the action~\eqref{eqn:vilaction} takes the form
\begin{equation}
    \label{eqn:wdwvile}
    (\nabla^2_0 - U_0 - H_q )\psi = 0,
\end{equation}
where the operator $H_0 = \nabla^2_0 - U_0$ is the part that survives when we neglect all the quantum variables $q^\nu$ and their conjugated momenta. The smallness of the quantum subsystem is ensured by the existence of a small parameter $\epsilon$ for which
\begin{equation}
    \label{eqn:epsilonvile}
    \frac{H_q \psi}{H_0 \psi}= O(\epsilon),
\end{equation}
where $\epsilon$ is a small parameter proportional to $\hbar$. The wave function of the Universe can be written as
\begin{equation}
    \label{eqn:wavevile}
    \psi(h,q) = A(h) e^{i S(h)}\chi(h,q).
\end{equation}
In order to perform a WKB expansion as a series in $\epsilon$ in a properly way, we point out that the potential term $U_{0}(h)$ is of the order $\epsilon^{-2}$ and the action $S(h)$ is of the order $\epsilon^{-1}$.
So, if we consider the wave funcion~\eqref{eqn:wavevile} inside the equation~\eqref{eqn:wdwvile} we obtain, at the lowest order in $\epsilon$, the Hamilton-Jacobi equation for S:
\begin{equation}
    \label{eqn:order0vile}
    g^{\alpha \beta} (\nabla_\alpha S)(\nabla_\beta S) + U = 0,
\end{equation}
and at the next order we obtain
\begin{equation}
\label{eqn:order1vile}
    2 \nabla A \nabla S + A \nabla^2 S + 2i \nabla S \nabla \chi - H_q \chi = 0
\end{equation}
The terms of the equation~\eqref{eqn:order1vile} can be decoupled in a pair of equation making use of the Adiabatic Approximation\cite{adiabatic}. It consist in requiring that the semiclassical evolution be principally contained in the semiclassical part of the wave function, while the quantum part depends on it only parametrically. The adiabatic approximation is therefore expressed by the condition
\begin{equation}
\label{eqn:adapprox}
|\partial_h A(h) | \gg | \partial_h \varphi (h,q) |
\end{equation}
Using the relation~\eqref{eqn:adapprox} in the~\eqref{eqn:order1vile} we obtain
\begin{equation}
\label{eqn:order1result}
\frac{1}{A}\nabla (A^2 \nabla S) = 0 \quad , \quad 2i \nabla S \nabla \chi - H_q \chi = 0
\end{equation}
The first equation represents the conservation of the current obtained neglecting the quantum part of the wave function (\ref{eqn:wavevile}). The second equation can be recasts in a Schr\"{o}dinger-like form for the quantum subsystem using the relation $\dot{h}^\alpha = 2 N \nabla^\alpha S$ in order to obtain
\begin{equation}
\label{eqn:schro1vile}
i \frac{\partial \chi}{\partial t} = N(t) H_q \chi
\end{equation}

The adiabatic scenario here described will be adopted in what follows to the Bianchi I model by considering the isotropic variable $a$ (\textit{i.e.} the Universe volume) as the quasi-classical component and the anisotropies $\beta_{\pm}$ like the pure quantum variables.

\section{Canonical Bianchi I Quantum Dynamics}
\label{sec:4}
In this section we consider a model in order to study the possibility to introduce a cyclical behaviour for the Bianchi I Universe in presence of a matter contribution. Let us consider a Universe described by a Bianchi I model filled with three different kinds of matter: stiff matter, a negative cosmological constant and an ultrarelativistic component. The description of the model will be done with respect to the Misner-like variables $\{ a, \beta_{\pm}\}$ and in this case the superHamiltonian constraint takes the form
\begin{equation}
\label{eqn:hbianchi1}
\mathcal{H} = \frac{G}{24 \pi a^3 c^3 } \biggl[ -a^2 p^2_a + p^2_{+} + p^2_{-} - \Lambda a^6 + \mu^2_{st} + \mu^2_{ur} a^2 \biggr ]
\end{equation}
where $\Lambda >0$ is the magnitude of the negative cosmological constant contribution and $\mu_{st}$, $\mu_{ur}$ represent respectively the stiff matter and the ultrarelativistic matter contributions. The canonical quantization of the model will be perform in the $p_a$-polarization. We will analyze a comparison between the case in which the polymer paradigm is not applied (the standard case) and the case when all the configuration space variables exhibit a discrete nature. Through this comparison we will see how the polymer quantization effects the nature of  the initial singularity.

\begin{figure}[h!]
\includegraphics[scale=.67]{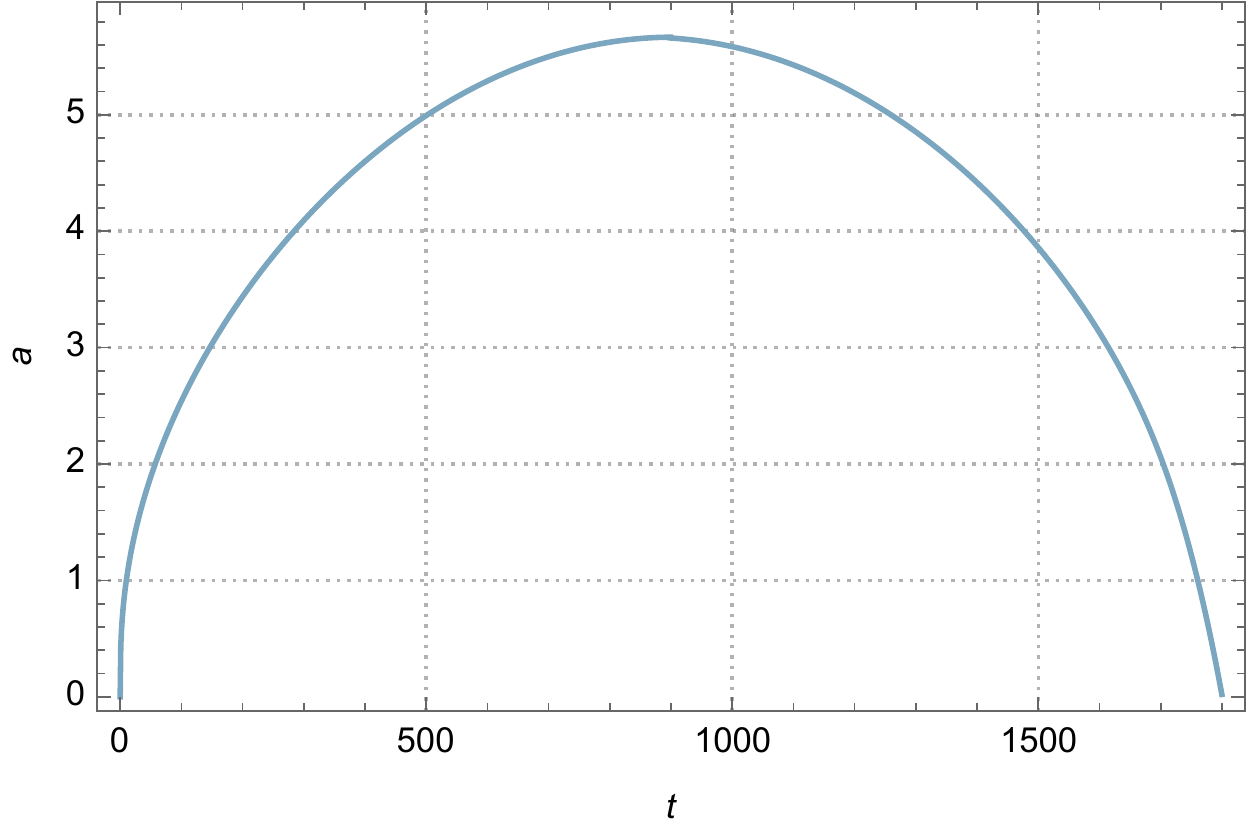}
\caption{The blue line represents the standard behavior $a(t)$ evaluated from the Eq.(\ref{eqn:ham_eq}). The consideration of the negative cosmological constant term is the reason of the presence of the two singularities, one in the past and one in the future. }
\label{fig:nopoly}
\end{figure}

We start studying a canonical quantization imposing that the physical states $\psi$ being annihilated by the operator $\mathcal{H}$, the quantum version of the superhamiltonian constraint~\eqref{eqn:hbianchi1}. We have that the action of the operator $\{ \hat{p}_a, \hat{p}_{+}, \hat{p}_{-} \}$ is multiplicative, while $\{ \hat{a}, \hat{\beta}_{+}, \hat{\beta}_{-} \}$ act as a derivative operators:
\begin{equation}
\label{eqn:operatoraction}
\hat{a} = i \hbar \frac{\partial}{\partial p_a} = i \hbar \partial_{p_a} \quad , \quad \hat{\beta}_{\pm} = i \hbar \frac{\partial}{\partial p_{\pm}} = i \hbar \partial_{p_{\pm}}
\end{equation}
The quantum counterpart of the superhamiltonian constraint~\eqref{eqn:hbianchi1}, the WDW equation, can be written as
\begin{equation}
\label{eqn:wdwbianchi1}\
\biggl[ \hbar^2 p^2_a \partial^2_{p_a} + p^2_{+} + p^2_{-} + \hbar^6 \Lambda \partial^6_{p_a}  + \mu^2_{st} - \hbar^2 \mu^2_{ur} \partial^2_{p_a} \biggr ] \psi = 0
\end{equation}
Following the Vilenkin interpretation of the wave function, we choose to assign the semiclassical character to the isotropic variable $a$, while the anisotropies $\{p_{+}, p_{-} \}$ characterize the quantum subsystem. After this, we can write the wave function of the Universe, following~\eqref{eqn:wavevile}, as
\begin{equation}
\label{eqn:wavebianchi}
\psi(p_a,p_{\pm}) = \chi(p_a) \varphi(p_a,p_{\pm}) = A(p_a)e^{\frac{i}{\hbar}S(p_a)}\varphi(p_a, p_{\pm})
\end{equation}
Considering for the wave function the shape above in the WDW equation leads, to the lowest order in $\hbar$, to the Hamilton Jacobi equation for the classical action
\begin{equation}
\label{eqn:HJnopoly}
p_a^2 = -\Lambda (\dot{S})^4 + \frac{\mu^2_{st}}{(\dot{S})^2} + \mu^2_{ur}
\end{equation}
where $(\dot{*}) \equiv \frac{\partial}{\partial p_a}$ and clearly the anisotropies or their conjugated momenta, which are the quantum variables, does not appear. Indeed, the lowest order in the WKB expansion performed respect to the $\hbar$ parameter takes into account the semiclassical behavior of the whole system, and this regard only the isotropic variable. Making a comparison between~\eqref{eqn:HJnopoly} and the superhamiltonian constraint~\eqref{eqn:hbianchi1} when the anisotropies are frozen (so that the term $\frac{p^2_{+} + p^2_{-}}{a^3}$ is neglected), we can establish the connection $\dot{S} = a$ and rewrite the equation~\eqref{eqn:HJnopoly} as
\begin{equation}
\label{eqn:HJnopoly2}
p_a^2 = -\Lambda a^4 + \frac{\mu^2_{st}}{a^2} + \mu^2_{ur}
\end{equation}
Then, it is possible to obtain the explicit solution for $a=a(t)$ making use of the equation~\eqref{eqn:HJnopoly2} together with the Hamiltonian Equation
\begin{equation}
\label{eqn:ham_eq}
\frac{da}{dt}= \frac{\partial \mathcal{H}}{\partial p_a} = -\frac{G p_a}{12 \pi c^3 a},
\end{equation}
in order to achieve
\begin{equation}
\label{eqn:ham_eq2}
\frac{da}{dt} = \frac{G}{12 \pi c^3} \frac{1}{a} \sqrt{-\Lambda a^4 + \frac{\mu^2_{st}}{a^2} + \mu^2_{ur}}.
\end{equation}
This differential equation is not analytically integrable, thus we must perform a numerical study It is possible to obtain an information about the maximum value of the  scale factor $a$ from the square root in the Eq.(\ref{eqn:ham_eq2}). Indeed, from $\frac{da}{dt}=0$ we obtain the condition ( $-\Lambda a^4 + \frac{\mu^2_{st}}{a^2} + \mu^2_{ur}) = 0$ that tells us
\begin{equation}
\label{eqn:amax_nopoly}
    a_{max}(t) = \sqrt{ \biggl( \frac{2}{3 B} \biggr ) ^{1/3} \mu_{ur}  + \frac{B}{18 \Lambda} }
\end{equation}
where $B=9 \Lambda^2 \mu_{st} + \sqrt{3} \sqrt{27 \Lambda^4 \mu^2_{st} -4 \Lambda^3 \mu^3_{ur}}$. 
It is easy to see, as reported in Fig.\ref{fig:nopoly}, that such a solution describe an evolution of the Universe of the scale factor between two singular points via a turning point ensured by the presence of the negative cosmological constant, see Fig 

The next order in the WKB expansion gives us an equation that by means of the adiabatic approximation can be decoupled in two distinct contributions. We require that the $p_a$-evolution is mainly contained in the amplitude $A$, while the isotropic variation of the quantum part $\varphi$ is negligible, so we can express che adiabatic condition as
\begin{equation}
\label{eqn:ad_approx}
|\partial_{p_a} \chi (p_a) | \gg |\partial_{p_a} \varphi (p_a, p_{\pm})|
\end{equation}
This provide with the two independent equation
\begin{gather}
\frac{i p^2_a}{A}(A^2 S^\prime)^\prime + \frac{3 i \Lambda}{A}(A^2 (S^\prime)^5 )^\prime - \frac{i \mu^2_{ur}}{A}(A^2 S^\prime)^\prime = 0 \\
i 2 \varphi ^\prime (S^\prime) p^2_a + 6 \Lambda i \varphi^\prime (S^\prime)^5 -2 i \mu^2_{ur} (S^\prime) + (p^2_{+} + p^2_{-} ) \varphi = 0
\end{gather}
The first equation corresponds to the conservation of the current $\nabla_{p_a} J^{p_a} = 0$, while the second equation provides the evolution of the quantum subsystem
\begin{equation}
\label{eqn:sch_nopoly}
2 i a \dot{\varphi} (p^2_a + 3 \Lambda a^4 - \mu^2_{ur}) = (p^2_{+} + p^2_{-} ) \varphi
\end{equation}
where $\dot{\varphi}= \frac{\partial \varphi}{\partial p_a}$. It is possible to write a Schr\"{o},inger-like equation for the quantum wave function $\varphi$ using the relation $\frac{\partial \varphi}{\partial p_a} = \frac{\partial \varphi}{\partial t} \frac{\partial t}{\partial p_a} = \frac{\partial \varphi}{\partial a} \frac{\dot{a}}{\dot{p_a}}$. We obtain $\dot{p_a}$ by differentiating with respect to the synchronous time the equation~\eqref{eqn:HJnopoly2}
\begin{equation}
\dot{p_a} = -2 \dot{a} \bigl [ 2 \Lambda a^3 + \mu^2_{st} a^{-3} \bigr ]
\end{equation}
and then we can rewrite the equation~\eqref{eqn:sch_nopoly} as
\begin{equation}
\label{eqn:schro_nopoly}
i \hbar \frac{\partial \varphi}{\partial \tau} = (p^2_{+} + p^2_{-}) \varphi
\end{equation}
where we used the relation
\begin{equation}
\label{eqn:tau_nopoly}
\tau = \int \frac{da}{2 a^2 \sqrt{-\Lambda a^4 + \frac{\mu^2_{st}}{a^2} +\mu^2_{ur}}}
\end{equation}
The solution of the Eq.~\eqref{eqn:schro_nopoly} corresponds to a wave function of the form
\begin{equation}
\label{eqn:wave_nopoly}
\varphi(\tau,p_{\pm}) = C e^{\frac{i}{\hbar}E \tau} e^{\frac{i}{\hbar}k_{+}p_{+}} e^{\frac{i}{\hbar}k_{-}p_{-}}
\end{equation}
with $E = \frac{(k^2_{+} + k^2_{-})}{\hbar^2}$and where $(k_{+}, k_{-})$ are the quantum numbers associated to the anisotropies, for which is valid the dispersion relation $k_{\pm} = \hbar <\beta_{\pm}>$. We now analyze the behavior of the quantum variables through the anisotropies. Let us introduce the Ehrenfest Theorem\cite{ehrenfest}, a useful instrument to study the evolution of a quantum operator. Let $\ket{\varphi}$ be the state of the quantum subsystem built starting by the wave function~\eqref{eqn:wave_nopoly}. The expectation value of the quantum operators $\{ \hat{\beta}_{+}, \hat{\beta}_{-} \}$ corresponds to
\begin{equation}
\braket{\hat{\beta}_{\pm}} = \braket{\varphi | \hat{\beta}_{\pm} | \varphi}
\end{equation}
The time derivative of the expectation value of a time independent operator $A$ is given by
\begin{equation}
\frac{d}{dt}\braket{A} = \frac{1}{i\hbar}\braket{\bigl [ A, H  \bigr ]}
\end{equation}
where $H$ is the Hamiltonian of the system. The only non vanishing position-momentum commutators are $\bigl[ \beta_{+} , p_{+} \bigr] = \bigl [ \beta_{-}, p_{-} \bigr] = i\hbar$, so that we can apply it to the anisotropies in order to obtain
\begin{equation}
\label{eqn:ehre_nopoly}
\frac{d\braket{\hat{\beta}_\pm}}{dt} = \frac{G}{24 i \hbar \pi a^3 c^3} \braket{ \bigl[ \beta_{\pm}, p^2_{\pm} \bigr]} = \frac{G \braket{p_{\pm}}}{12 \pi a^3 c^3}
\end{equation}
where we used the commutation rule $[A, BC] = [A,B]C + B[A,C]$. Applying the Ehrenfest theorem to the operators $\hat{p}_{\pm}$, it is possible to show that them expectation values are constants of motion:
\begin{equation}
\frac{d \braket{\hat{p}_{\pm}}}{dt} = \frac{1}{i \hbar} \braket{ \bigl[ p_{\pm}, H \bigr]} = 0 \to \braket{\hat{p}_{\pm}} = const.
\end{equation}
Using the above result in the equation~\eqref{eqn:ehre_nopoly} with the time-evolution for the isotropic variable in the equation~\eqref{eqn:ham_eq2} we arrive at the differential equation
\begin{equation}
\label{eqn:ehre_np}
\frac{d\braket{\hat{\beta}_{\pm}}}{d a}= \frac{G \braket{\hat{p}_{\pm}}}{c^3 a^2\sqrt{-\Lambda a^4 + \frac{\mu^2_{st}}{a^2} + \mu^2_{ur}}},
 \end{equation}
that has no analytic solution and has to be numerical integrated. 

We can also study the average values of the anisotropies by analyzing the numerical evolution of the wave packets built as follows 
\begin{equation}
\label{eqn:packetwave_nopoly}
\Psi_{k^*_{\pm}}(\tau(a),p_{\pm}) = \int \int dk_{\pm} e^{-\frac{(k_{+}-k^*_{+})^2}{2 \sigma^2_{+}}} e^{-\frac{(k_{-}-k^*_{-})^2}{2 \sigma^2_{-}}} \varphi(\tau, p_{\pm}),
\end{equation}
where we choose Gaussian weights to peak the wave packets.

\begin{figure}[h!]
\includegraphics[scale=.67]{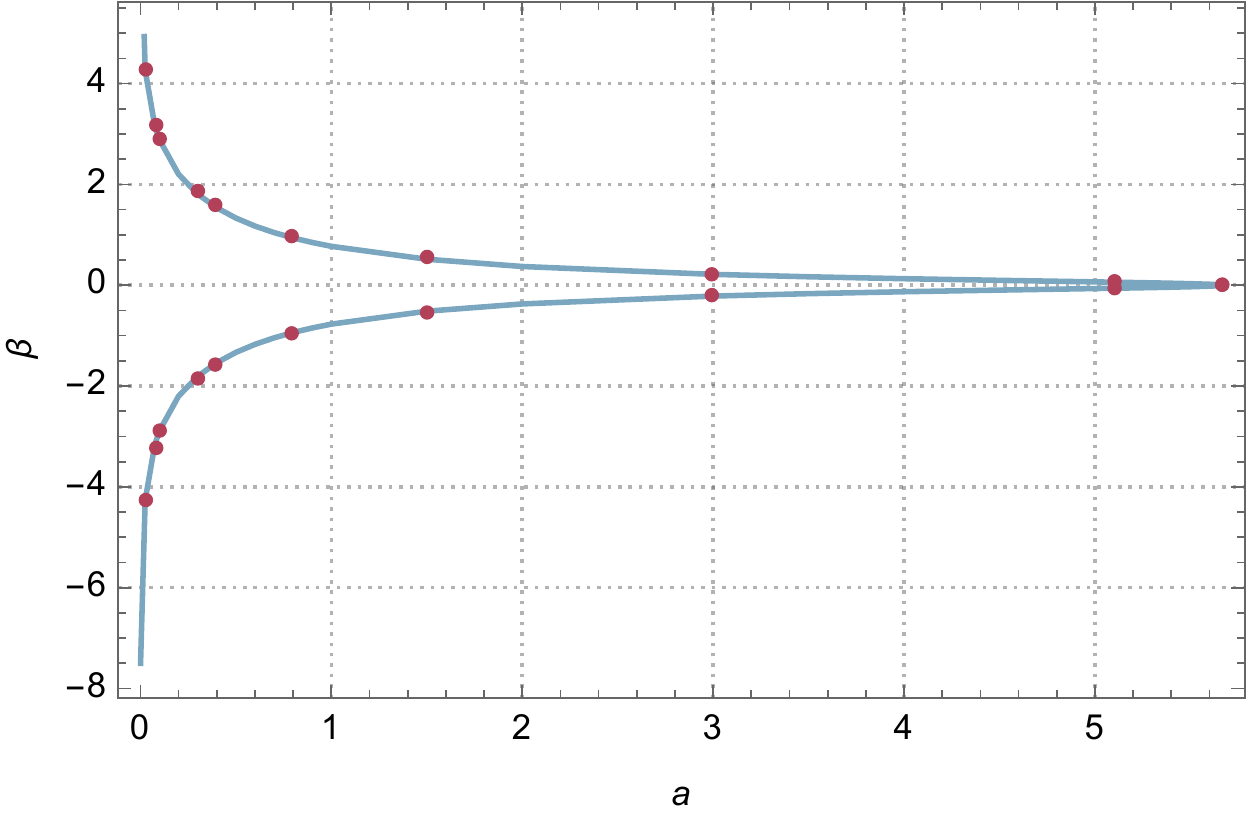}
\caption{The red points represent the position of the maximum of the wave packet $|\Psi_{k_{\pm}^{*}}(\tau(a),\beta_{\pm})|$
evaluated via numerical integration. The continuous blue line represents the trajectory evaluated with the same parameters from the Ehrenfest theorem. In both cases the analysis confirm the diverging behavior of the anisotropies next to the singularities}
\label{fig2}
\end{figure}

Actually, the numerical integration of the maximum of the wave packets and of the Ehrenfest theorem provides the same dynamical behavior of the Universe anisotropies, see FIG \ref{fig2}. It is worth stressing that in the singularities ($a \simeq 0$) the average values of the anisotropies diverge.

\section{Polymer Bianchi I Quantum Dynamics}
\label{sec:5}
As shown before for the one-dimensional particle case, if we consider the position variable $q$ as a discrete variable, it is impossible to associate a differential quantum operator for the conjugated momenta $p$. Thus, through the polymer procedure we can identify an approximate version of the operator $\hat{p}$ which, in the momentum polarization, acts multiplicatively on the states of the system. Following equation~\eqref{eqn:p2approx} we consider the polymer paradigm substitution for the whole configuration space variables:
\begin{equation}
\label{eqn:p2_approx}
p^2_i \to \frac{2 \hbar^2}{\lambda^2} \biggl [ 1 - \cos{ \biggl ( \frac{\hbar p_i}{\lambda} \biggr )} \biggr],
\end{equation}
where $i = \{ a, \beta_{+}, \beta_{-} \}$.
This way, the superhamiltonian constraint~\eqref{eqn:hbianchi1} and the WDW equation~\eqref{eqn:wdwbianchi1} are modified as

\newpage
\begin{widetext}
\begin{gather}
\label{eqn:H_poly}
\mathcal{H} = \frac{G}{24 \pi a^3 c^3 } 
\biggl[ -a^2 \frac{2 \hbar^2}{\lambda^2} \biggl [ 1 - \cos{ \biggl ( \frac{\hbar p_a}{\lambda} \biggr )} \biggr]
+ \frac{2 \hbar^2}{\lambda^2} \biggl [ 1 - \cos{ \biggl ( \frac{\hbar p_{+}}{\lambda} \biggr )} \biggr]
+\frac{2 \hbar^2}{\lambda^2} \biggl [ 1 - \cos{ \biggl ( \frac{\hbar p_{-}}{\lambda} \biggr )} \biggr]
- \Lambda a^6 + \mu^2_{st} + \mu^2_{ur} a^2 \biggr ]=0 \\
\label{eqn:WDW_poly}
\biggl[ \hbar^2
\frac{2 \hbar^2}{\lambda^2} \biggl [ 1 - \cos{ \biggl ( \frac{\hbar p_a}{\lambda} \biggr )} \biggr]
\partial^2_{p_a} 
+ \frac{2 \hbar^2}{\lambda^2} \biggl [ 1 - \cos{ \biggl ( \frac{\hbar p_{+}}{\lambda} \biggr )} \biggr]
+ \frac{2 \hbar^2}{\lambda^2} \biggl [ 1 - \cos{ \biggl ( \frac{\hbar p_{-}}{\lambda} \biggr )} \biggr]
+ \hbar^6 \Lambda \partial^6_{p_a}  + \mu^2_{st} - \hbar^2 \mu^2_{ur} \partial^2_{p_a} \biggr ] \psi = 0
\end{gather}
\end{widetext}
As in Sec.\ref{sec:4}, we now consider the wave function of the Universe~\eqref{eqn:wavebianchi} inside the~\eqref{eqn:WDW_poly}. At the lowest order of the expression in $\hbar$ we obtain the semiclassical level and the Hamilton-Jacobi equation obtained can be written as

\begin{equation}
\label{eqn:HJ_poly}
p_a = \frac{\hbar}{\lambda} \arccos{ \biggl [ 1 - \frac{\lambda^2}{2}\biggl( -\Lambda a^4 + \frac{\mu^2_{st}}{a^2} + \mu^2_{ur} \biggl) \biggr ]},
\end{equation}
where we have identified again $\dot{S} = a$. From the superhamiltonian~\eqref{eqn:H_poly} we can write the Hamilton equation for the isotropic variable

\begin{equation}
\label{eqn:ham_eq_poly}
\dot{a} = \frac{da}{dt} = \frac{\partial H}{\partial p_a} = \frac{G}{12 \pi \lambda c^3 a}\sin{ \biggl ( \frac{\lambda p_a}{\hbar} \biggr )}
\end{equation}
Introducing the Eq.~\eqref{eqn:HJ_poly} in the Eq.~\eqref{eqn:ham_eq_poly} and making use of the trigonometric relation $\sin{\arccos{x}} = \sqrt{1 - x^2}$, we arrive to

\begin{widetext}
\begin{equation}
\label{eqn:dadt_poly}
\frac{da}{dt} = \frac{G}{12 \pi c^3 a} \sqrt{\biggl ( -\Lambda a^4 + \frac{\mu^2_{st}}{a^2} + \mu^2_{ur} \biggr ) \biggl [ 1 - \frac{\lambda^2}{4} \biggl ( -\Lambda a^4 + \frac{\mu^2_{st}}{a^2} + \mu^2_{ur}  \biggr ) \biggr ]}
\end{equation}
\end{widetext}
Looking at the latter equation it is immediate to show the existence of two particular values where $\frac{da}{dt}$ changes the sign and represent respectively maximum value and minimum value for the scale factor $a$

\begin{gather}
\label{eqn:a_min_poly}
a_{min} = \sqrt{-4 \biggl ( \frac{2}{B} \biggr )^{1/3} +\lambda^2 \mu^2_{ur} \biggl ( \frac{2}{B} \biggr )^{1/3} + \frac{1}{3 \lambda^2 \Lambda} \biggl ( \frac{B}{2} \biggr )^{1/3}} \\
a_{max} = \sqrt{\mu^2_{ur} \biggl (\frac{2}{3C} \biggr )^{1/3} +  \biggl ( \frac{C}{18 \Lambda^3} \biggr )^{1/3}},
\label{eqn:a_max_poly}
\end{gather}
with 
\begin{equation}
B=27 \lambda^6 \Lambda^2 \mu^2{st} + \sqrt{729 \lambda^{12} \Lambda^4 \mu^4_{st} ( 12 \lambda^2 \Lambda - 3 \lambda^4 \Lambda \mu^2_{ur})^3}
\end{equation}
\begin{equation}
C= 9 \Lambda^2 \mu^2_{st} + \sqrt{81 \Lambda^4 \mu^4_{st} - 12 \Lambda^3 \mu^6_{ur}}.
\end{equation}

Again, it is not possible to obtain an analytic solution for the~\eqref{eqn:dadt_poly}, so we operate a numerical integration starting from the maximum value of the scale factor $a$ to the minimum and compare the result with what we obtain studying~\eqref{eqn:ham_eq2}.
  \begin{figure}
{\includegraphics[scale=0.6]{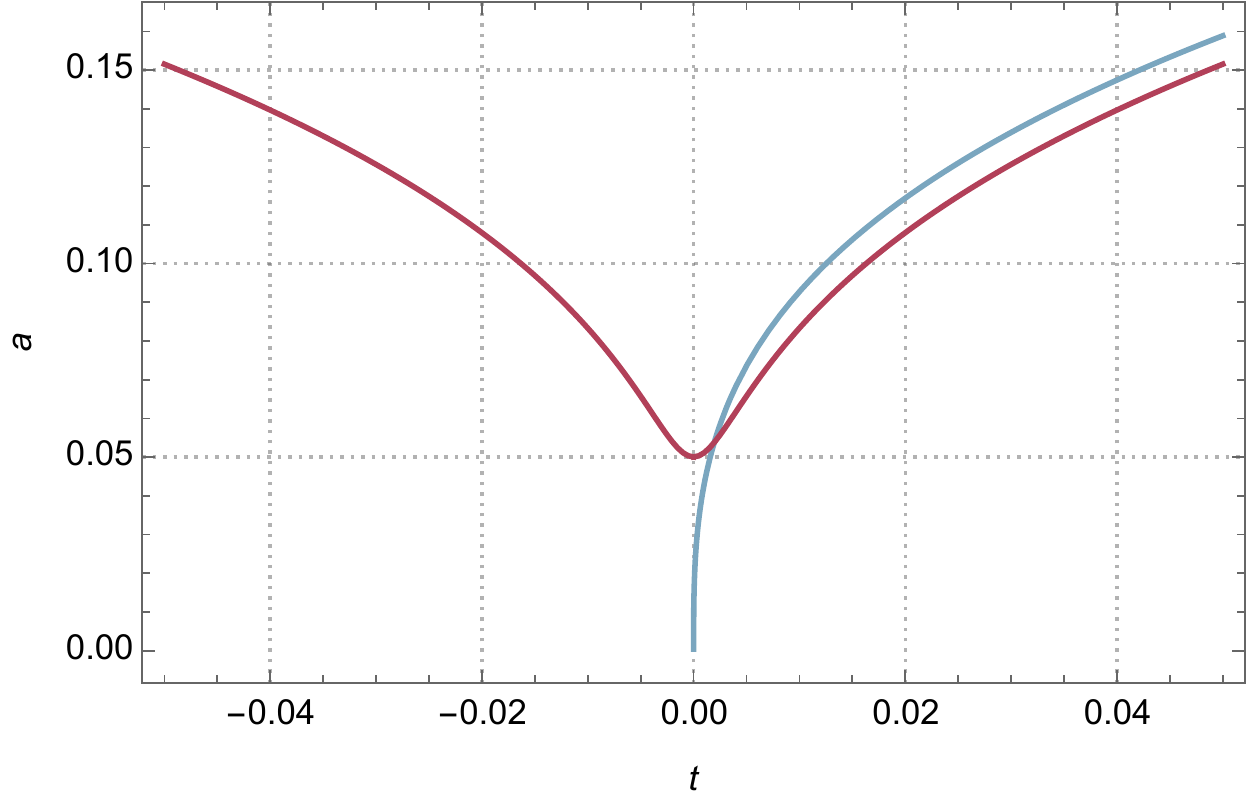}} \\
{\includegraphics[scale=0.6]{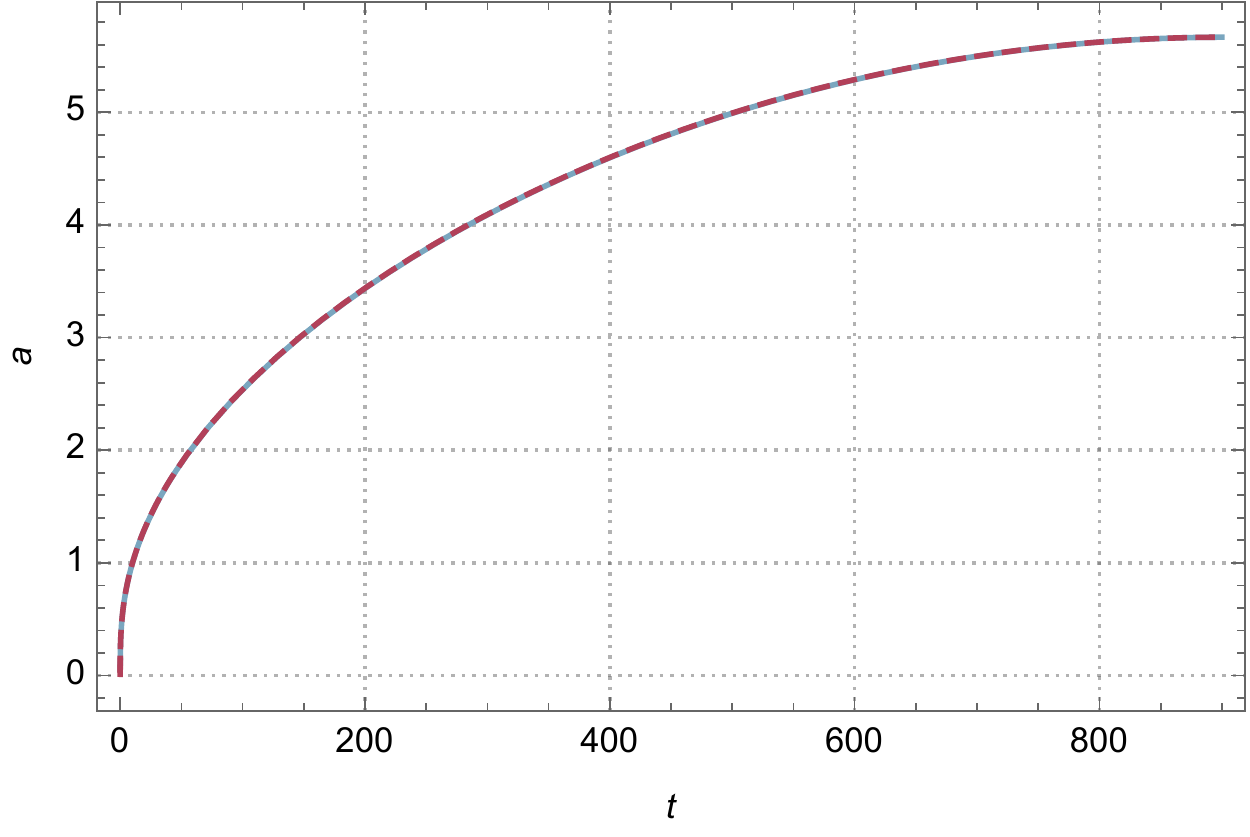}}
\caption{The behaviour of the scale factor $a(t)$ near and away from the singularity when the polymer modification is
considered. For late times, the evolution of the isotropic variable resembles the same standard solution shown in FIG.\ref{fig2} while, for early times, the polymer substitution implies the replacement of the Big Bang singularity with a Big bounce.}
\label{fig:a3m}
\end{figure}
What we see from FIG.~\ref{fig:a3m} is that far from the singularity the standard and the polymer solution for the scale factor tend to become identical. Instead, for $t\to0$, we can see that the polymer contribution becomes predominant so that the scale factor $a$ reaches the value given by~\eqref{eqn:a_min_poly} and not $a=0$. Furthermore,  joining together the route associated to the expanding Universe and the one to the collapsing we get a cyclical behavior of the isotropic variable which oscillate between the big bounce (induced by the polymer cut-off physics) and the turning point (due to the small negative cosmological constant). 

Proceeding in the same way as in the previous Section, we have at the first order in $\hbar$ the quantum part of the wave function $\varphi$.

\begin{equation}
\label{eqn:quantum_poly}
2 i \hbar \dot{\varphi} \biggl ( \frac{a}{\lambda^2} \biggl[ 1 - \cos{(\lambda p_a)} \biggr ] + 3\Lambda^5 - \mu^2_{ur} a \biggr ) = p^2_{\pm} \varphi
\end{equation}
where $\dot{\varphi}=\frac{\partial \varphi}{\partial p_a} = \frac{\partial \varphi}{\partial a}\frac{\dot{a}}{\dot{p_a}}$. We get $\dot{p_a}$ through a differentiation of the relation~\eqref{eqn:HJ_poly} and the equation \eqref{eqn:dadt_poly} in order to write

\begin{equation}
\label{eqn:dot_p}
\dot{p_a} = \frac{(2 \Lambda a^4 + \mu^2_{st} a^{-2}}{12 \pi a^2}
\end{equation}
This way we can write the equation~\eqref{eqn:quantum_poly} as

\begin{equation}
\label{eqn:sch_poly}
i \hbar \frac{\partial \varphi}{\partial \tau} = (\frac{2 \hbar^2}{\lambda^2} \biggl [ 1 - \cos{ \biggl ( \frac{\hbar p_{+}}{\lambda} \biggr )} \biggr] + \frac{2 \hbar^2}{\lambda^2} \biggl [ 1 - \cos{ \biggl ( \frac{\hbar p_{-}}{\lambda} \biggr )} \biggr]) \varphi
\end{equation}
with a time-like variable $\tau$

\begin{equation}
\label{eqn:tau_poly}
\tau = \int \frac{da}{2 a^2 \sqrt{\bigl ( -\Lambda a^4 + \frac{\mu^2_{st}}{a^2} + \mu^2_{ur} \bigr ) \bigl [ 1 - \frac{\lambda^2}{4} \bigl ( -\Lambda a^4 + \frac{\mu^2_{st}}{a^2} + \mu^2_{ur}  \bigr ) \bigr ]}}
\end{equation}
We can see that if we implement the limit $\lambda \to 0$ we turn back to a time-like variable $\tau$ like the one obtained in~\eqref{eqn:tau_nopoly}. This allow us to write down the analytic version of the quantum part of the wave function $\varphi$

\begin{equation}
\label{eqn:wave_poly}
\varphi(\tau,p_{\pm}) = C e^{\frac{i}{\hbar}E \tau} e^{\frac{i}{\hbar}P^{pol}_{+}p_{+}} e^{\frac{i}{\hbar}P^{pol}_{-}p_{-}},
\end{equation}
where

\begin{gather}
E = \frac{2}{\lambda^2} \biggl[ 2 - \cos{\lambda p_{+}} - \cos{\lambda p_{-}} \biggr] \\
P^{pol}_{\pm} = \frac{1}{\lambda}\arccos{\biggl [ ( 1 - \frac{k^2_{\pm} \lambda^2}{2}) \biggr]}
\end{gather}
and $\{k_{+}, k_{-} \}$ are the quantum numbers associated to the anisotropies. We now analyze the behavior of the quantum variables associated to the Universe anisotropies. We will use again the Ehrenfest Theorem and its application to $\hat{\beta}_{\pm}$. We thus have

\begin{widetext}
\begin{equation}
\label{eqn:ehre_poly}
\frac{d\braket{\hat{\beta}_\pm}}{dt} = \frac{1}{i \hbar} \braket{ \bigl[ \beta_{\pm} , H \bigr]} = \frac{G}{24 i \hbar \pi a^3 c^3} \braket{ \bigl[ \beta_{\pm}, p^2_{\pm} \bigr]} = \frac{G \braket{\sin{(\lambda p_{\pm})}}}{12 \pi a^3 c^3}
\end{equation}
\end{widetext}

As for the time-like variable $\tau$, we can not get an analytic result for the Ehrenfest Theorem, so we perform a numerical integration in order to obtain
\begin{widetext}
\begin{equation}
\label{eqn:ehren_poly}
\braket{\beta}_{\pm} = \int \frac{da G \braket{\sin{(\lambda p_{\pm})}}}{c^3 a^2 \sqrt{\bigl ( -\Lambda a^4 + \frac{\mu^2_{st}}{a^2} + \mu^2_{ur} \bigr ) \bigl [ 1 - \frac{\lambda^2}{4} \bigl ( -\Lambda a^4 + \frac{\mu^2_{st}}{a^2} + \mu^2_{ur}  \bigr ) \bigr ]}}
\end{equation}
\end{widetext}

\begin{figure}
    \includegraphics[scale=0.65]{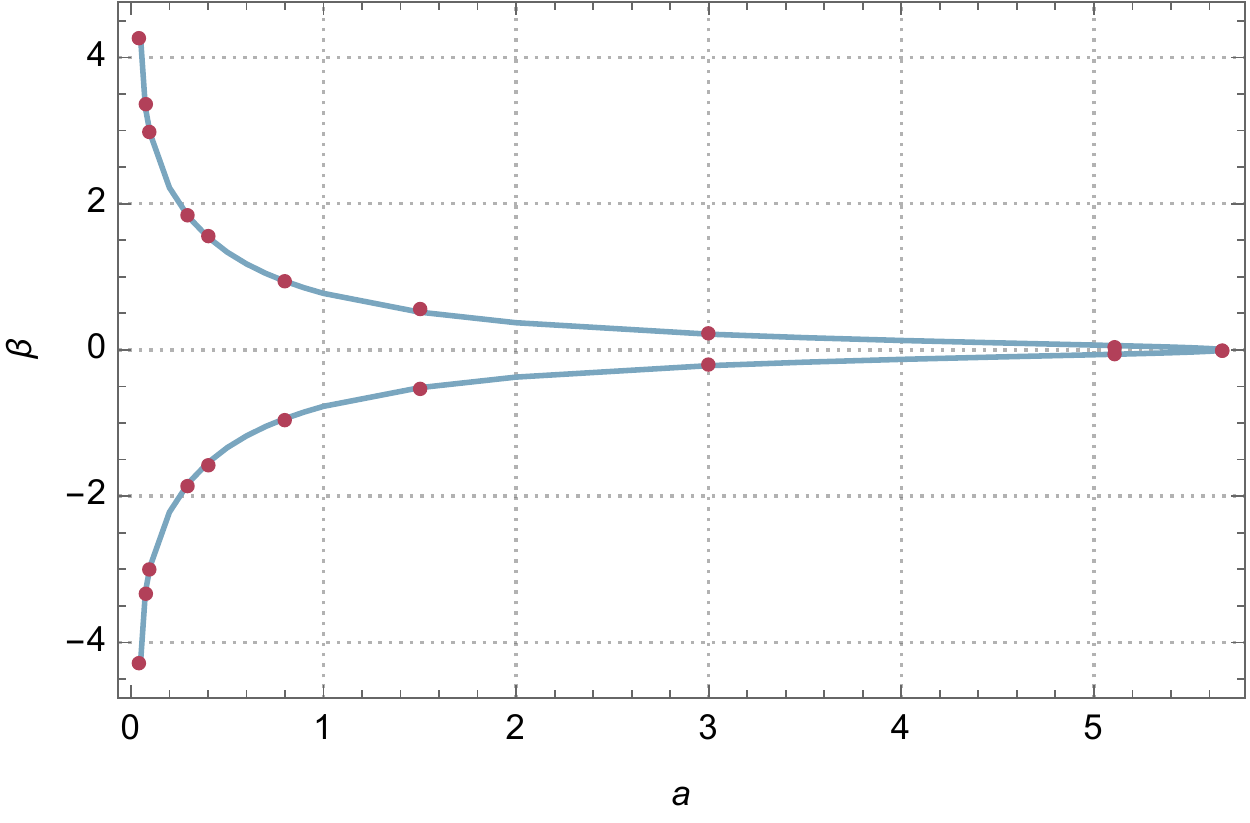}
    \caption{The behaviour of $\beta(a)$. The blue line represent the result of the numerical integration of~\eqref{eqn:ehren_poly}, while the red dots are the results of the numerical integration of the wave packet obtained from~\eqref{eqn:wavepacket_poly}. With the introduction of the polymer application, the singularity is removed and the anisotropies don't diverge.}
   \label{fig:packet_poly}
   \end{figure} 
   
In FIG.~\ref{fig:packet_poly} is shown the behaviour of the quantum expectation value $\hat{\beta}_{\pm}$ in the polymer case compared to the standard dynamics. We can easily see that far away from singularity the two trajectories are almost overlapped, and the polymer contributions became important only when $a \to a_{min}$, since the singularity is removed due to the semiclassical behaviour of the isotropic variable in this scenario. An additional confirm of the dynamics of the anisotropies can be provide by studying the behavior of the maximum of the wave packet built from the wave function~\eqref{eqn:wave_poly} as follows

\begin{equation}
\label{eqn:wavepacket_poly}
\Psi_{k^*_{\pm}}(\tau,p_{\pm}) = \int \int dk_{\pm} e^{-\frac{(k_{+}-k^*_{+})^2}{2 \sigma^2_{+}}} e^{-\frac{(k_{-}-k^*_{-})^2}{2 \sigma^2_{-}}} \varphi(\tau, p_{\pm})
\end{equation}
where we choose Gaussian weights to peak the wave packets. The evolution of the wave packets has been studied through a numerical integration and as we can see in the FIG.~\ref{fig:packet_poly}, the position of the maximum of the wave packet $\Psi_{k^*_{+}} (\tau, p_{\pm}) $ as a function of $a$ overlaps exactly the trajectory of the anisotropies obtained by the Ehrenfest theorem in~\eqref{eqn:ehren_poly}. Thus, we can conclude that the anisotropy dynamics is significantly affected by the cut-off physics, especially because of the scale factor no longer goes to $0$. Indeed, the minimum values of the variables $\{ \beta_{+}, \beta_{-} \}$ remain finite during the whole evolution of the cyclical Universe. This feature has a relevant physical meaning in view of implementing a Big-Bounce cosmology\cite{brandenberger}
for the actual Universe. Since we can control the value taken by the anisotropies variables across the bounce, via a suitable choice of the initial conditions, then it is possible to argue that the Universe remain sufficiently close to an isotropic configuration, even during the Planck era.

\section{Phenomenological considerations}
We want now to face some phenomenological questions concerning the physical meaning and the
cosmological implementation of the considered model.

Actually, our model seems to be peculiar for two reason: the choice of the cosmological model and the
matter sources. Nevertheless, especially for what concerns the behavior of the
primordial Universe near the cosmological singularity, this is actually far from the truth.

The choice of stiff and ultrarelativistic matter contributions is well-posed near the initial
singularity, since they represent the energy contributions due to the kinetic term of the inflaton
field and due to the very hot thermal bath of ultrarelativistic particles(see respectively
\cite{primordial},\cite{kolb}). In particular, the presence of the stiff matter is justified by its functional
behavior as the inverse squared volume of its energy density: independently on the initial
conditions, such a contribution is destinated to dominate near enough to the initial singularity,
corresponding to vanishing Universe volume (we observe that the equation of state of stiff matter
is $p = \rho$ and that $p > \rho$ would corresponds to a superluminar sound velocity). 
The Bianchi I model is associated to zero spatial curvature and this feature makes it less
general than other Bianchi types, especially the Bianchi IX one (the most general, together the
type VIII, allowed by the homogeneity constraint and having the closed Robertson-Walker geometry
as isotropic limit \cite{landau},\cite{gravitation}). 

However, as it is well known \cite{scalarfield},\cite{berger} ,
in the presence of the stiff matter contribution (that mimics the kinetic term of a scalar field), the Bianchi IX Universe becomes chaos-free and admits a stable Kasner regime (a Bianchi I type dynamics) towards the initial singularity. 
Furthermore, such a cosmological paradigm can be easily extended to the inhomogeneous sector
\cite{BKL82},\cite{scalarfield}\cite{montani2000}; infact, near enough to the initial singularity,
the space points dynamically
decouple and locally the homogeneous behavior qualitatively holds \cite{primordial}.
These considerations led us to consider the Bianchi I dynamics in the context of thepolymer quantum
cosmology, as the properties of the singularity are investigated. Nonetheless, some non-trivial
questions remains open. Indeed, it is not clear if the scenario traced above remains valid when
the Bianchi IX dynamics with stiff matter is analyzed in the polymer semi-classical approximation.
Not only the presence of the chaotic behavior, but also the existence itself of the singularity must be addressed
by a detailed specific study, whose issues could also depend on the particular representation we refer to. 
A qualitative but solid argument is offered by the observation that, even in the polymer semi-classical dynamics, the Bianchi IX model must be represented by a piecewise evolution, whose single step is a potential-free Bianchi I regime (the form of the Bianchi IX potential and its vanishing behavior in an increasing triangular region of the $\beta_{\pm}$-plane are substantially not affected by the polymer representation). 

We can use the mean values we got form the Ehrenfest theorem to estimate the asymptotic behavior of the potential next to the singular point: we expect that the finiteness of the anisotropies mean values due to the existence of a bounce, can mitigate the role played by such a spatial curvature contribution. 

For this reason, considering now the Bianchi IX potential $V_{IX}(\beta_{\pm})$, were $V_{IX}$ explicitly reads as
\begin{multline}
\label{eqn:vbianchi_ix}
V(\beta) = \frac{1}{3}e^{-8 \beta_{+}}-\frac{4}{3}e^{-2\beta_{+}} \cosh{2 \sqrt{3}\beta_{-}} \\
+1 +\frac{2}{3}e^{4 \beta_{+}} (\cosh{4 \sqrt{3}\beta_{-}} -1)
\end{multline}

\begin{figure}[h!]
\includegraphics[scale=.67]{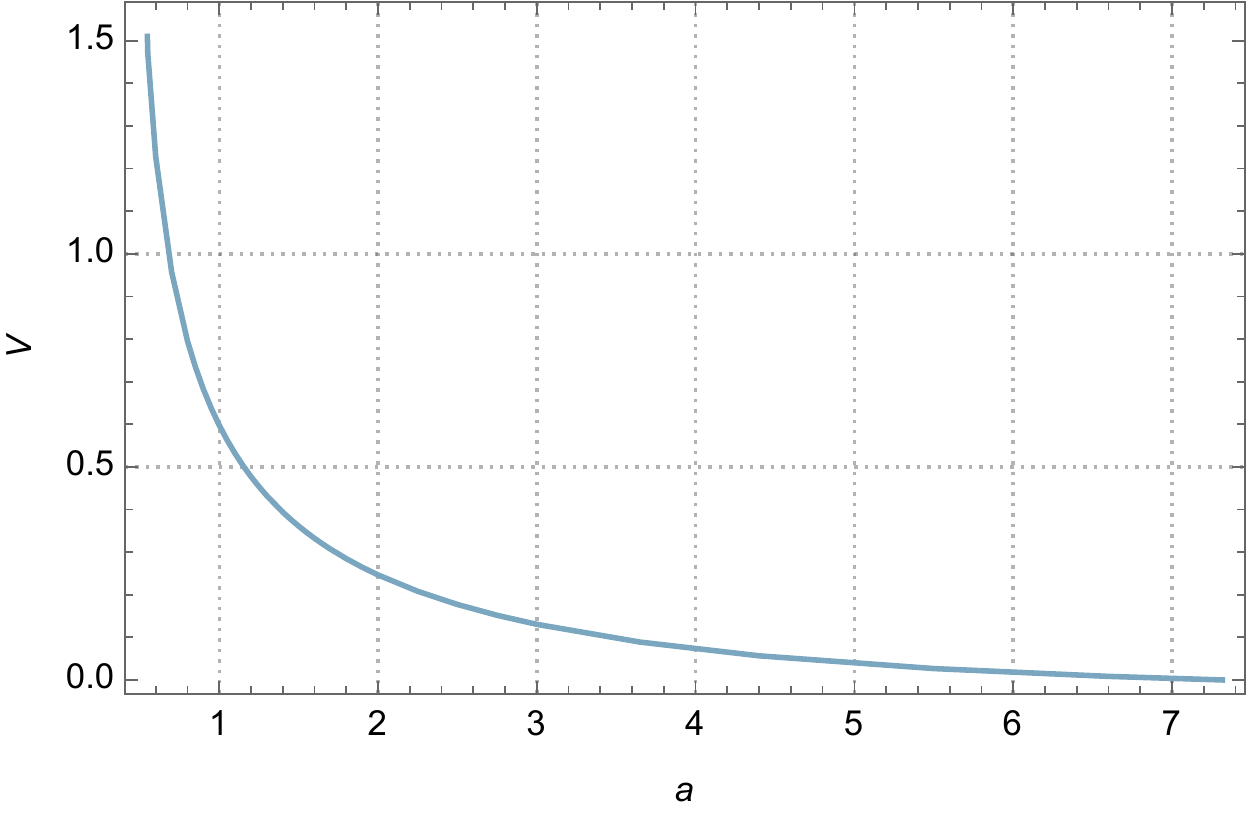}
\caption{The behaviour of  $V_{IX}(\langle \beta_{\pm}(a) \rangle)$.
The blue line represent the result obtained for~\eqref{eqn:vbianchi_ix} using
the numerical integration of \eqref{eqn:ehren_poly} with similar terms. With the polymer application,
the curvature potential term does not exhibits the typical diverging behavior next to the singularity but rather it remains finite crossing the bounce.}
\label{fig:vbianchi9}
\end{figure}

In Fig.\ref{fig:vbianchi9}, we plot the quantity $V_{IX}(\langle \beta_{\pm}(a) \rangle)$. We use, to evaluate the potential term, the expectation values of the anisotropies as a function of the isotropic variable $a$, as expressed by the integral representation \eqref{eqn:ehren_poly}, in order to study the evolution of $V_{IX}$ towards the dingularity. As it is possible to see in the Figure, for suitable choice of the free parameters (i.e. of the initial conditions for the model),
it reaches a maximum (no longer diverging) value in the Big-Bounce. 
By other words, differently from the ordinary classical Bianchi IX dynamics, the existence of a
cut-off on the vanishing behavior of the spatial volume induces a finiteness of both the
anisotropies and the potential term (i.e. the contribution of the model spatial curvature).
As a result, we can infer that the information we get on the Big-Bounce in the present Bianchi I
analysis will remain almost valid even when the most general Bianchi IX case is considered and,
in this respect,
the inclusion of stiff matter in the dynamics appears a privileged choice both from a physical
(it represent the kinetic term of the inflaton field) and a dynamical (it limits the role of
spatial curvature in Einsteinian dynamics) point of view. 

The situation is a bit more subtle for what concerns the late evolution of the considered model with respect to a Bianchi IX cosmology. In fact, in the latter, the re-collapse of the Universe is due to its non-zero spatial curvature, exactly as it takes place for a closed Robertson-Walker geometry. Instead, in the considered scenario, the existence of a turning point for the Bianchi I dynamics (and hence the emergence of a cyclical Universe) is associated with the presence of a
small negative cosmological constant. Although our dynamical paradigm is simpler than a more
realistic Bianchi IX model, it captures the proper features of the re-collapse mechanism (in the
isotropic limit the spatial curvature behaves like $a$, while the negative cosmological
term like $a^3$), but avoiding the really non-trivial questions concerning the isotropization
process of the Bianchi IX Universe, for instance via the inflationary paradigm (see the general
analysis in \cite{kirillovmontani2002}). 

The only significant limitation of the considered scenario is that it could not be reconciliated
with a dark energy associated to a positive cosmological constant. In fact, since our negative
cosmological term will cause a turning point in the future of the actual Universe, it would be
cancelled by a positive cosmological constant accelerating the Universe today. 
However, for any other dark energy equation of state $p = w\rho$ with $-1 < w
<-1/3$, such a 
contribution would decay as the Universe expands and the negative cosmological
constant soon or later, would dominate, producing the Universe recollapse. 
This difficulty would be naturally overcome in the case of a Bianchi IX cosmology and it does not
affect the main focus of the present analysis, i.e. the emergence of a non-singular Universe, 
having finite anisotropies, as result of the semi-classical polymer dynamics. However, about the
role that a positive cosmological constant could play in the context of a cyclical Universe,
especially in view of its capability to stop the associated oscillations, see the study in
\cite{referee}.

\section{Conclusions}

In this paper, we analyzed a Bianchi I cosmology in the presence of stiff
matter, ultrarelativistic matter and a small negative cosmological constant.
The main aim of the present analysis was to demonstrate that such a model constitute 
a good paradigm for a cyclical Universe, whose anisotropy degree of freedom are 
always finite and, in principle, they can be controlled via suitable initial conditions. 

In order to achieve the scenario of a cyclical Universe, we first implement the 
polymer quantization procedure to all the minisuperspace variables and then we consider 
an adiabatic quasi-classical separation of the Universe volume (i.e. the
isotropic part of the cosmic scale factors) with respect to the pure quantum dynamics 
of the anisotropy variables (i.e. the real gravitational degrees of freedom, here
described via the Misner variables $\beta_{\pm}$). 

We demonstrated how the quasi-classical 
evolution of the Universe volume is characterized by the emergence of
a Big-Bounce cosmology, associated to a minimum of the corresponding
configurational variable $a$, due to the cut-off physics implied 
by the polymer approach. 
The presence of a later turning point in the Bianchi I dynamics is ensured as effect of a 
small negative cosmological constant. In this respect, a
phenomenological consideration is needed to better focus our dynamical paradigm. 

Such a constant must ensure a re-collapsing dynamics in the future of the
present Universe and therefore its value 
is postulated, much smaller (at least $0.1$ time) than the positive
cosmological constant that it is expected to accelerate the actual Universe. 

It is worth noting that, in the more general case of a Bianchi IX model, the
role here played by the negative cosmological constant would be replaced by the spatial
curvature of the Universe, responsible for a potential term in the Hamiltonian
dynamics. 
In this respect, the present study must be 
regarded as the prototype of the more general analysis of a Bianchi IX
cosmology, see \cite{moriconiBB2}, which nonetheless would contain additional 
subtleties about the behavior of the emerging potential in the adiabatic
quasiclassical approximation. 

Finally, we stress how the finiteness of the 
anisotropy degrees of freedom during the evolution of the present model 
is a relevant cosmological issue. 

In fact, the non-divergent character of the average value of the variables 
$\beta_{\pm}$ allows to infer that, under a suitable restriction of the
initial conditions, the Universe can remain mainly isotropic across the Big-Bounce. 

This result opens a new scenario about the interpretation of the present
Universe as a step in the cyclical dynamics for a singularity-free cosmology. 
This new point of view could also affect the 
understanding of the basic paradoxes which led to the construction of an
inflation paradigm \cite{primordial}.

\end{document}